\documentclass{ws-procs975x65}

\begin{document}

\title{Nonsingular Black Holes in Palatini Extensions of General Relativity}

\author{Gonzalo J. Olmo}
\address{Departamento de F\'{i}sica Te\'{o}rica and IFIC,
Centro Mixto Universidad de Valencia - CSIC. Universidad de
Valencia, Burjassot-46100, Valencia, Spain \\ E-mail: {gonzalo.olmo@csic.es}}

\author{D. Rubiera-Garcia}
\address{Departamento de F\'{i}sica, Universidad de Oviedo,
Avenida Calvo Sotelo 18, 33007, Oviedo, Asturias, Spain \\ E-mail: {rubieradiego@gmail.com}}

\begin{abstract}
We discuss static, spherically symmetric solutions with an electric field in a quadratic extension of general relativity formulated in the Palatini approach (assuming that metric and connection are independent fields). Unlike the usual metric formulation of this theory, the field equations are second-order and ghost-free. It is found that the resulting black holes present a central core whose area is proportional to the Planck area times the number of charges. Some of these solutions are nonsingular. In this case, the charge-to-mass ratio implies that the core matter density is independent of the specific amounts of charge and mass and of order the Planck density.
\end{abstract}


\bodymatter

\section{Introduction}\label{sec:Intro}
In order to gain some insights on how the internal structure of black holes could get modified by quantum gravitational effects, one can consider quadratic corrections to the dynamics of general relativity (GR), which arise when the quantum renormalizability of the matter fields is considered in curved space\cite{parker-toms,birrell-davies} and also when GR is seen as an effective theory\cite{cembranos}. Here we address this problem assuming that the space-time possesses independent metric and affine structures (Palatini formalism). Though quadratic gravity theories have been studied in the literature in a variety of situations, the Palatini formulation has comparatively received much less attention. Our motivation to consider the Palatini approach\cite{Olmo2011a} comes from recent results that support this framework as a very promising arena to explore aspects of quantum gravity phenomenology\cite{Olmo2011b,Olmo-Singh, Bounces}.

\section{Theory and solutions.}
Our model is described by the following Einstein-Maxwell action with Planck-scale corrections in the gravitational sector ($l_P=\sqrt{\hbar G/c^3} \equiv$ Planck length)
\begin{eqnarray}\label{eq:action}
S&=&\hbar \int \frac{d^4x \sqrt{-g}}{16\pi l_P^2}\left[R+l_P^2\left( a R^2+ R_{(\mu\nu)}R^{(\mu\nu)}\right)\right]- \frac{1}{16\pi} \int d^4x \sqrt{-g} F_{\alpha\beta}F^{\alpha\beta} \nonumber \ .
\end{eqnarray}
Taking independent variations of (\ref{eq:action}) with respect to metric and connection yields
\begin{eqnarray}
f_R R_{\mu\nu}-\frac{1}{2}f g_{\mu\nu}+2f_Q R_{\mu\alpha}{R^{\alpha}}_{\nu}&=&\kappa^2 T_{\mu\nu} \label{eq:f(R,Q)-metric}\\
\nabla_\alpha\left[\sqrt{-g}\left(f_R g^{\beta\gamma}+2f_Q R^{\beta\gamma}\right)\right]&=&0 \ , \label{eq:f(R,Q)-connection}
\end{eqnarray}
where $\kappa^2=8\pi l_P^2/ \hbar$, $f=R+l_P^2\left(a R^2+ R_{(\mu\nu)}R^{(\mu\nu)}\right)$, $f_X\equiv \partial_X f$, $Q\equiv R_{(\mu\nu)}R^{(\mu\nu)}$, ${T_\mu}^\nu=\frac{1}{4\pi}\left[{F_{\mu}}^{\alpha} {F^{\nu}}_{\alpha}-\frac{F_{\alpha\beta}F^{\alpha\beta}}{4}\delta_\mu^\nu\right]$ and $F_{\mu\nu}=\partial_{\mu}A_{\nu}-\partial_{\nu}A_{\mu}$ is the field strength tensor of the vector potential $A_{\mu}$. Through algebraic manipulations, one can show that the symmetric part of the connection $\Gamma_{\alpha\beta}^\gamma$ can be written as the Levi-Civita connection of an auxiliary metric $h_{\mu\nu}$ related with  $g_{\mu\nu}$ as follows
\begin{equation}\label{eq:hmn-general}
{h}_{\mu\nu}=\sqrt{\det \Sigma}{\left({\Sigma^{-1}}\right)_\mu}^\alpha g_{\alpha\nu} \ , \ {h}^{\mu\nu}=\frac{g^{\mu\alpha}{\Sigma_\alpha}^\nu}{\sqrt{\det \Sigma}}
\end{equation}
where ${\Sigma_\alpha}^\nu=f_R\delta_\alpha^\nu+2f_Q {P_\alpha}^\nu$ is a function of ${T_\mu}^\nu$ and ${P_\alpha}^\nu\equiv g^{\mu\nu}R_{\alpha\mu}$. The antisymmetric part of the connection is set to zero for simplicity. For a spherical electric field, the field equations can be written as [$\sigma_\pm\equiv \left(1\pm {l_P^2r_q^2}/{r^4}\right)$ and $r_q^2=\kappa^2q^2/4\pi$]
\begin{equation}
{R_\mu}^\nu(h)=\frac{r_q^2}{2r^4}\left(\begin{array}{ccc}
-\frac{1}{\sigma_+} \hat{I}& \hat{0} \\
\hat{0} & \frac{1}{\sigma_-}\hat{I}
\end{array} \right)  \label{eq:Ricci-h4} \ ,
\end{equation}
which smoothly recover GR in the limit $l_P\to 0$. Using a spherically symmetric line element for $h_{\mu\nu}$, one can show that the metric $g_{\mu\nu}$ can be expressed as\cite{OR12c}
\begin{equation}\label{eq:g}
g_{tt}=-\frac{A(z)}{\sigma_+} \ , \ g_{rr}=\frac{\sigma_+}{\sigma_-A(z)}  \ , \
 A(z)=1-\frac{\left[1+\delta_1 G(z)\right]}{\delta_2 z \sigma_-^{1/2}}  \ ,
\end{equation}
where $z\equiv r/\sqrt{r_q l_P}$, $\delta_1=\frac{1}{2r_S}\sqrt{\frac{r_q^3}{l_P}}$,  $\delta_2= \frac{\sqrt{r_q l_P} }{r_S}$, and $r_S\equiv 2M$ represents the Schwarzschild radius of the vacuum solution. All the information about the geometry is encapsulated in $G(z)$, which satisfies $\frac{dG}{dz}=\frac{z^4+1}{z^4\sqrt{z^4-1}}$. For $z \gg 1$ we have $G(z)\approx -1/z-3/10z^5$, which leads to $A(r)\approx 1-r_S/r+r_q^2/2r^2-r_S r_q^2l_P^2/2r^5$ and recovers GR when $r\gg l_P$. The general solution is given by
\begin{equation}\label{eq:G-hyper}
G(z)=\beta +\frac{1}{2} \sqrt{z^4-1} \left[f_{\frac{3}{4}}(z)+f_{\frac{7}{4}}(z)\right] \ ,
\end{equation}
where $f_{\lambda}(z)= {_{2}F}_1[\frac{1}{2},\lambda,\frac{3}{2},1-z^4]$, and $\beta\approx -1.74804$ is a constant resulting from matching the $z \gg 1$ and $z\to 1$ expansions. This constant is very important in this theory. In fact, when the charge-to-mass ratio $\delta_1$ is set to the value $\delta_1^*=-1/\beta$, the resulting geometry is smooth everywhere, as can be seen from the line element
\begin{equation}
ds^2\approx \frac{1}{2}\left(1-{\delta_1^*}/{\delta _2 }\right)\left[-dt^2+(dr^*)^2\right]+(r_q l_P) d\Omega^2 \
\end{equation}
and from the direct computation of curvature scalars such as $R$, $R_{\mu\nu}R^{\mu\nu}$ and $R_{\alpha\beta\gamma\delta}R^{\alpha\beta\gamma\delta}$. For any other value of $\delta_1$ curvature divergences arise at $z=1$.

\section{Physical properties}
Expressing the charge as $q=N_q e$, where $e$ is the electron charge and $N_q$ the number of charges, we can write $r_q=\sqrt{2\alpha_{em}}N_q l_P$, where $\alpha_{em}$ is the fine structure constant. With this we find that the area of the $z=1$ surface is given by $A_{core}= N_q\sqrt{2\alpha_{em}} A_P$,  where $A_P=4\pi l_P^2$ is Planck's area. This suggests that each charge sourcing the electric field has associated an elementary quantum of area of magnitude $\sqrt{2\alpha_{em}} A_P$. From this it follows that the ratio of the total charge $q$ by the area of this surface gives a universal constant, $\rho_q=q/(4\pi r_{core}^2)=(4\pi\sqrt{2})^{-1}\sqrt{c^7/(\hbar G^2)}$, which up to a factor $\sqrt{2}$ coincides with the Planck surface charge density. Furthermore, the regularity condition $\delta_1=\delta_1^*$ sets the following mass-to-charge relation $\frac{M}{(r_ql_P)^{3/2}}=\frac{1}{4\delta_1^*}\frac{m_P}{l_P^3}$,
which indicates that the matter density inside a sphere of radius $r_{core}$ is another universal constant, $\rho_{core}^*=M/V_{core}=\rho_P/4\delta_1^*$,  independent of $q$ and $M$.
From the large $z$ expansion, one can verify that the location of the external horizon of the nonsingular charged black holes is essentially the same as in GR, $r_+= r_S\left(1+\sqrt{1-4\delta_1^*/(N_q\sqrt{2\alpha_{em}})}\right)/2$. For a solar mass black hole, the number of charges needed to avoid the $z=1$ singularity  is just $N_\odot=(2r_S \delta_1^* /l_P)^{2/3}/\sqrt{2\alpha_{em}}\approx 2.91 \times 10^{26}$, which is a tiny amount on astrophysical terms (in fact, the Sun contains $\sim 10^{54}$ protons). In general, $N_q=N_\odot (M/M_\odot)^{2/3}$ implies that in astrophysical scenarios $r_+\approx r_S$.

\section{Conclusions and perspectives}

We have shown with an exactly solvable model that nonperturbative effects at the Planck scale can generate important modifications on the innermost structure of black holes while keeping their macroscopic features essentially unchanged. This is an aspect that has been recursively used in heuristic discussions in the literature but without solid mathematical support. Our model, formulated \`{a} la Palatini, provides an explicit realization of this idea. A deeper understanding of the properties of the solutions obtained here could shed useful new light on the properties of black hole singularities, the mechanisms that prevent them, and the fate of black hole evaporation\cite{MartinezAsencio:2012xn}. Investigations of these issues are currently underway.

\section{Acknowledgments}
This work has been supported by the Spanish grant FIS2011-29813-C02-02 and the JAE-doc program of the Spanish Research Council (CSIC).

\end{document}